# Using Mobile Phone Sensor Technology for Mental Health Research: Integrated Analysis to Identify Hidden Challenges and Potential Solutions

Tjeerd W. Boonstra[1, *], Jennifer Nicholas[1], Quincy J. J. Wong[1,2], Frances Shaw[1], Samuel Townsend[1], Helen Christensen[1]

[1]Black Dog Institute, UNSW Sydney, Sydney, Australia
[2]Western Sydney University, Sydney, Australia

[*]**Corresponding author**
Tjeerd Boonstra
Black Dog Institute
Hospital Rd
Randwick, NSW, 2031
Australia
+61 2 9382 9285
t.boonstra@unsw.edu.au

## Abstract

**Background**: Mobile phone sensor technology has great potential in providing behavioral markers of mental health. However, this promise has not yet been brought to fruition.
**Objective:** The objective of our study was to examine challenges involved in developing an app to extract behavioral markers of mental health from passive sensor data.

**Methods**: Both technical challenges and acceptability of passive data collection for mental health research were assessed based on literature review and results obtained from a feasibility study. Socialise, a mobile phone app developed at the Black Dog Institute, was used to collect sensor data (Bluetooth, global positioning system, and battery status) and investigate views and experiences of a group of people with lived experience of mental health challenges (N=32).

**Results**: On average, sensor data were obtained for 55% (Android) and 45% (iPhone OS) of scheduled scans. Battery life was reduced from 21.3 hours to 18.8 hours when scanning every 5 minutes with a reduction of 2.5 hours or 12%. Despite this relatively small reduction, most participants reported that the app had a noticeable effect on their battery life. In addition to battery life, the purpose of data collection, trust in the organization that collects data, and perceived impact on privacy were identified as main factors for acceptability.

**Conclusions**: Based on the findings of the feasibility study and literature review, we recommend a commitment to open science and transparent reporting and stronger partnerships and communication with users. Sensing technology has the potential to greatly enhance the delivery and impact of mental health care. Realizing this requires all aspects of mobile phone sensor technology to be rigorously assessed.

# Introduction

## Background

Mobile phone sensor technology has great potential in mental health research, providing the capability to collect objective data on behavioral indicators independent of user input [1-3]. With the plethora of sensors built into mobile phones, passive collection of a wide range of behavioral data are now possible using the device most people carry in their pockets [4]. Passive data collection operates in the background (requires no input from users) and allows measurement of variables longitudinally with detailed moment-to-moment information and collection of temporal information on dynamic variables, such as users' feelings and activity levels. Given that these digital records reflect the lived experiences of people in their natural environments, this technology may enable the development of precise and temporally dynamic behavioral phenotypes and markers to diagnose and treat mental illnesses [5].

An ever-growing number of mobile phone apps have been developed to passively collect sensor data for mental health purposes, for example, Purple Robot is a mobile phone sensor data acquisition platform developed at Northwestern University that is available on Android devices. The platform gives access to a range of sensors including device information, battery level, Bluetooth and Wi-Fi, global positioning system (GPS), accelerometer, and communication logs [6] and has been used in research studies on mental health in the general community [7,8]. The Beiwe Research Platform was developed at Harvard University to collect research-quality data from mobile phone sensors, including use patterns, on both Android and iPhone OS (iOS) platforms in primarily clinical samples. The app collects a range of sensor data including that obtained from GPS, accelerometer, communication logs, Wi-Fi and Bluetooth (Android only), and battery use [9]. Another notable example is the large-scale Copenhagen Networks Study, a research project studying social networks in 1000 university students, which provided Android mobile phones to collect Bluetooth, Wi-Fi, communication logs, and GPS sensor data [10]. These different software and methodological approaches have resulted in varied behavioral indicators being targeted, different features extracted, and different statistical approaches used to link behavioral data to mental health.

## Behavioral Markers of Mental Health

Depression is associated with a number of behavioral changes, of which sociability and activity are the most studied using mobile phone sensor data. Social connectedness is a key factor in mental health and well-being. Social isolation, perceptions of loneliness, lower perceived social support, and fewer close relationships have all been linked to depression [11,12]. Mental health is also affected by the location of individuals within their social network and the number and strength of their social connections [13]. Traditionally, social networks have been investigated using self-reported data, the reliability of which has been questioned [14]. Sensor-enabled mobile phones allow for the collection of passive data to map social networks of proximity using sensor data including that obtained from Bluetooth. Eagle et al [15] were able to differentiate friends from nonfriends accurately using temporal and spatial patterns of Bluetooth data. As far as we know, mobile phones that collect Bluetooth sensor data have not yet been used in mental health studies. However, Pachucki et al [16] have used wearable proximity sensors to map a social network in high school children, showing that adolescent girls with more depressive symptoms have smaller social networks.

Depression is also associated with decreased activity and motivation and increased sedentary behavior [17]. Cross-sectional data indicates that people with depression are less likely to be active than people without depression [18]. Furthermore, longitudinal studies have shown that baseline depression is associated with increased sedentary behavior over time [18] and that low physical activity at baseline is associated with increased depression [19]. Again, mobile phone sensors, particularly GPS, are well placed to monitor an individual's location, physical activity, and movement. Initial research in a small sample (N=18) has indicated potential features of GPS data, such as a lower diversity of visited places (location variance), more time spent in fewer locations, and a weaker 24-hour, or circadian, rhythm in location changes, that are associated with more severe depression symptoms [7].

## Challenges of Mobile Phone Sensor Technology

Despite the potential of mobile phone sensor technology in mental health research, this promise has not yet been brought to fruition. The use of mobile phone sensor technology for mental health research poses several key challenges, both technical and issues specific to mental health apps. A primary technical challenge is the reliable collection of sensor data across mobile platforms and devices, for example, location data may be missing due to sensor failure to obtain GPS coordinates [20,21], participants not charging or turning off their phones, or unavailability of any network connections for a long period of time, hampering data transfer to servers [7,10]. The mode of data collection also influences data completeness, which can differ between operating systems. Passive collection of sensor data are easier to support on Android than iOS; about twice as many apps are available for Android than for iOS [22]. This likely reflects greater restrictions that iOS places on accessing system data and background activity, making personal sensing using iOS devices challenging.

Another technical issue is battery life. Frequent sampling of sensor data can consume a significant proportion of a mobile phone's battery [23]. Ultimately, if an app collecting sensor data are too resource-intensive, users' motivation to continue using it decreases [24], which may lead to the app being uninstalled, ceasing the flow of data to researchers. Optimizing passive data collection to obtain the most detailed information possible should therefore be balanced with expectations of users regarding battery consumption. This is a significant practical challenge faced by mobile sensing apps.

In addition, there are specific challenges for using mobile phone sensor technology for mental health purposes, such as the engagement and retention of users [25]. Increasingly, a user-centered design approach is considered an integral part of any mental health app development [26-29]. Individuals with the target disorder can provide important information about the direction and focus of the app as well as how they engage with an app given their symptom profile. For example, focus groups of individuals with Post-Traumatic Stress Disorder (PTSD) indicated that PTSD Coach was particularly useful for managing acute PTSD symptoms and helping with sleep [30]. Clinicians, on the other hand, can provide input into the design and functionality of an app from a therapeutic perspective. For example, clinicians indicated that an app for individuals with bipolar disorder to self-manage their symptoms should focus on medication adherence, maintaining a stable sleep pattern, and staying physically and socially active [31]. Codesign of mental health apps with end users and other stakeholders increases the likelihood that the app will be perceived as attractive, usable, and helpful by the target population [24]. Although design and usability issues are often discussed for apps that require active user engagement, it is also important for passive data collection apps to increase user engagement and retention because this will ensure lower rates of missing data and dropouts. Furthermore, many apps have an ecological momentary assessment (EMA) component to complement passive sensor data collection.

User perceptions of an app's confidential handling and use of data, as well as privacy and anonymity, are additional challenges of passive data collection [9,32,33]. Mental health data are highly sensitive because of the potential negative implications of unwanted disclosure [34]; therefore, uncertainty about whether a service is confidential can be a barrier to care [35]. Indeed, data privacy and confidentiality are major concerns for the users of mental health apps [36,37], but no consensus has yet been reached on ethical considerations that need to be addressed for the collection of passive sensor data. Moreover, user perceptions of security and privacy may differ; for example, Android and iOS users differ in characteristics such as age and gender [38] and also in their awareness about security and privacy risks of apps [39]. Deidentification may be used to the protect privacy of individuals [40] but may also remove information that is important to maintain the usefulness of data, depending on context and purpose for use [41]. Systems making use of predictive analysis techniques not only collect data but also *create* information about personal mental health status, for example, through identification of markers for risk [42]. Therefore, social impact needs to be considered beyond individual privacy concerns.

## Outline

In this study, we examined challenges of using mobile phone sensor technology for mental health research by analyzing results of a feasibility study that was conducted to test an app collecting passive sensor data. We analyzed the amount of sensor data that was collected, assessed the ability to quantify behavioral markers from Bluetooth and GPS data collected in a real-world setting, quantified battery consumption of the app, and examined user feedback on usability. No mental health questionnaires were administered as part of the feasibility study, although demographic and diagnostic data were available from the volunteer research register from which participants were drawn. We also investigated views of participants about acceptability of passive data collection for mental health research. The purpose of collecting this information was to build greater understanding of how social norms and perceptions around technology and data collection impact the feasibility, ethics, and acceptability of these technologies. We related results from our feasibility study to existing literature in these areas to identify common challenges of using mobile phone sensor technology in mental health research. We also drew some distinctions between available apps and made brief recommendations for the field going forward.

## Methods

### Mobile phone app

Socialise, a mobile phone app developed at the Black Dog Institute, was used to assess the feasibility and challenges of passive data collection in a group of volunteers. We developed Socialise as a native app in Java for Android and Objective-C for iOS to collect passive data (Bluetooth and GPS) and EMA. Building on the results of a previous validation and feasibility study [43,44], we implemented several changes to improve scanning rates on iOS and here we tested Socialise version v0.2. We used silent push notifications to trigger Bluetooth and GPS scans and to upload data to the server. Silent push notifications, along with the "content-available" background update parameter, were used to deliver a payload containing an operation code corresponding to either a Bluetooth or GPS scan or one of a number of data uploads. The allowable background time for processing a push notification is sufficient to perform these scans and record data, and we hence used silent push notification to overcome some of the limitations imposed by iOS on apps running in the background. In addition, we used the significant-change location service to improve data collection rates. Unlike Android devices, no mechanism exists on iOS to allow the app to relaunch when a device restarts. By subscribing to the significant-change location service, the app is notified when the device restarts and triggers a local notification reminding participants to resume data collection.

### Participants and Procedure

This study was approved by the University of New South Wales Human Research Ethics Committee (HC17203). Participants were recruited through advertisements disseminated through the Black Dog Institute volunteer research register. Individuals sign up on this register to volunteer for research. As part of the sign-up process, individuals provide demographics and diagnostic information (ie, mental disorders they have experienced in their lifetimes). To be able to participate in this study, individuals had to be 18 years or older, reside in Australia, speak English, and have a mobile phone running Android version 4.4 or newer or running iOS8 or newer. Interested individuals received a link to the study website where they could read participant information and provide consent. Of the 32 participants who provided consent to participate in the study, 31 also agreed to have their data made available on a public repository. Once they gave consent, participants received a link to install the Socialise app and a unique participant code. When participants opened the app, they were asked to give permission for the app to receive push notifications and collect location and Bluetooth data. Participants then had to fill in the unique participant code. Once the app opened, participants were asked to complete an entry survey, which included questions about the age of their mobile phone, the amount of time spent on their phone each day, and evaluation of their satisfaction with the onboarding process.

Participants were instructed to use the Socialise app for 4 weeks. Bluetooth and GPS data were collected during scans that were conducted at intervals of 8, 5, 4, or 3 minutes (equivalent to 7.5, 12, 15, and 20 scans per hour, respectively). Each scanning rate was tested for 1 week, and participants were instructed to use their phones normally for the duration of the study.

### Data Collection

We used the *BluetoothManager* private API on iOS devices to collect Bluetooth data, because the public *CoreBluetooth* API contains only functions for interacting with low-energy devices. It is currently not feasible to use Bluetooth Low Energy to map social networks in iOS [45]. To collect GPS data, the *CoreLocation* framework was utilized on iOS. The Android implementation leveraged the built-in Bluetooth APIs and *LocationManager* to collect Bluetooth and GPS data. Data acquisition settings were identical on iOS and Android, and both were set to collect Bluetooth, GPS, and battery data every 3, 4, 5, and 8 minutes.

Because the Bluetooth media access control address of a device is potentially personally identifiable information, these data were cryptographically hashed on the handset to ensure the privacy of participants. Hashing generates a consistent "signature" for each data item that cannot be reversed to reveal the original data value. To record only other mobile phones, detected devices were filtered according to the Bluetooth Core Specification. This involved removing any devices not matching the Class of Device 0×200 during the Bluetooth scan.

Participants were asked to complete a short questionnaire at the end of each week to document any problems that they encountered using the app. It included questions about whether they had changed phone settings (eg, turned off GPS or mobile data or turned on airplane mode), whether they used Bluetooth on their phone, and whether they thought the Socialise app impacted battery life. These findings were evaluated using a 7-point Likert scale. In addition, a set of questions about the acceptability of sensor data collection and some contextual information about that acceptability was collected at the end of the study.

### Data Analysis

Data completeness was assessed by comparing the number of Bluetooth and GPS scans that were scheduled for the duration of the study (9156 samples per participant) with the number of data samples that were uploaded by the app; that is, we scheduled scans every 3, 4, 5, and 8 minutes, each for a week (4 weeks), which comes to $20 \times 24 \times 7 + 15 \times 24 \times 7 + 12 \times 24 \times 7 + 7.5 \times 24 \times 7 = 9156$ total scans.

Most research using mobile phone Bluetooth to track social interactions has been performed in closed social networks [10,15,43,46]. In contrast, in this study, sensor data were collected from participants living in Australia who were unlikely to have social connections with each other. We therefore followed procedures described by Do et al [47] for analyzing Bluetooth data in a real-world setting. Instead of using Bluetooth to assess social connection between participants, Bluetooth was used to make a coarse estimate of human density around the user, which provides a rough proxy for social context. We first distinguished between known and unknown devices. Known devices were defined as devices that had been observed on at least 3 different days during the duration of the study. We then computed the average number of known and unknown devices that were detected at each hour of the day to obtain a social context profile for each participant.

We followed procedures outlined in Saeb et al [7] for analyzing GPS data. To identify location clusters, we first determined whether each GPS location data sample came from a stationary or a transition state. We calculated the time derivate to estimate movement speed for each sample and used a threshold of 1 km/h to define the boundary between the two states. We then used *K*-mean clustering to partition data samples in the stationary state into K clusters such that overall distances of data points to centers of their clusters were minimized. We increased the number of estimated clusters from 1 until the distance of the farthest point in each cluster to its cluster center fell below 500 m. We also estimated circadian movement, a feature that strongly correlated with self-reported depressive symptom severity [7]. Circadian movement measures to what extent participants' sequence of locations follows a 24-hour rhythm. To calculate circadian movement, we used least squares spectral analysis [48] to obtain the spectrum of GPS location data and estimate the amount of energy that fell with the 24-hour frequency bin. Circadian movement was then defined as the logarithm of the sum of energy for longitude and latitude [7].

The battery consumption of the Socialise app was estimated by varying the scanning rate each week. Varying scan rates enabled us to differentiate the battery consumption of the Socialise app from that of other apps running on the participants' mobile phones. We estimated the battery consumption of the Socialise app using linear regression, assuming that battery consumption scaled linearly with the number of scans performed per hour. To estimate battery consumption, we first extracted data samples when the battery was discharging and then computed the change in battery charge between scans. We next estimated the length of time for the battery to be exhausted separately for each scanning rate and device. We used a robust fitting algorithm, that is, reweighted least squares with the bisquare weighting function [49], to estimate the average battery consumption across devices and how it changed with scanning rate.

All analyses were performed using Matlab version R2018a (The MathWorks Inc, Natick, MA, USA) and the Matlab scripts used to analyze data are available at Zenodo: http://doi.org/10.5281/zenodo.1238408.

To evaluate user perceptions of battery consumption of the app, we compared responses on perceived impact on battery life across the 4 weeks of the study to assess whether perceived impact was affected by the actual scanning rate. To examine views of participants about the acceptability of passive data collection for mental health research, we compared their responses for different data types and contexts using a one-way repeated-measures one-way analysis of variance (ANOVA). Statistical analyses were performed using JASP version 0.8.3.1 (University of Amsterdam, the Netherlands). We also collected open responses to these questions, allowing for qualitative analysis. However, owing to the small number of responses, coding to saturation was not possible and we conducted a thematic analysis instead, dividing responses into categories to determine their approximate range.

## Results

### Participant Characteristics

Overall, 53 people expressed interest in participating in the study. Of these, 41 completed registration and gave informed consent. Of the 41, 1 participant was not eligible because the person did not live in Australia, 1 participant withdrew, 2 participants were unable to install the app on their mobile phones, and 5 participants did not respond to the follow-up email. The remaining 32 participants successfully installed the app on their mobile phones.

The age of participants was broadly distributed with the majority aged from 55 to 64 years (see Table 1). Most were female (23/30, 77%) and reported that they had been diagnosed with a mental disorder (23/32, 72%); depression and anxiety disorders were most commonly reported (Table 1). Participants reported using their mobile phones regularly, and most devices were less than a year old (15/30, 50%).

**Table 1.** Participant demographics.

| Characteristics | n | Percentage |
|---|---|---|
| **Sex (n=30)** | | |
| Male | 7 | 23 |
| Female | 23 | 77 |
| **Age in years (n=30)** | | |
| 18-24 | 5 | 17 |
| 25-34 | 6 | 20 |
| 35-44 | 5 | 17 |
| 45-54 | 4 | 13 |
| 55-64 | 7 | 23 |
| 65+ | 3 | 10 |
| **Mental disorder diagnosis (n=32)** | 23 | 72 |
| Depression | 22 | 69 |
| Bipolar disorder | 9 | 28 |
| Anxiety disorder | 17 | 53 |
| Schizophrenia | 0 | 0 |
| Personality disorder | 2 | 6 |
| Substance use disorder | 5 | 16 |
| Eating disorder | 7 | 22 |
| Autism spectrum disorder | 1 | 3 |
| Post-Traumatic Stress Disorder | 2 | 6 |
| Attention deficit hyperactivity disorder | 1 | 3 |
| **Daily phone usage (n=30)** | | |
| Less than 30 min | 2 | 7 |
| 30 min–1 h | 7 | 23 |
| 1–2 h | 4 | 13 |
| 2–3 h | 6 | 20 |
| More than 3 h | 11 | 37 |

### Data Completeness

Over the course of the study, 1 participant withdrew and another stopped participating. We therefore obtained sensor data from 28 of the 41 who consented to participate with a retention rate of 68%. Survey data were collected from 23 participants (participants who provided at least one response on the short questionnaire at the end of each week) and 13 participants completed the exit survey, as seen in Figure 1. Over the 4 weeks, a total of 9156 data points was scheduled for each participant. On average, 55 (19)% of scheduled samples were collected on Android and 45 (20)% on iOS, as seen in Figure 2. The figure shows the percentage of the number of scheduled samples (9156 samples per participant) that were collected on the devices used in the study. The x-axis lists the mobile phone model that each participant used.

The scanning rates did not significantly differ between operating systems ($t_{26}=-1.33$, $P=.19$, $d=0.53$). However, the number of scans that were collected varied considerably between devices (range 16.3%-95.4%), approximating normal distribution (iOS: $W=0.93$, $P=.20$; Android: $W=0.95$, $P=.65$). We also recorded the model of the device, but there did not appear to be a clear relationship with the scanning rate, as seen in Figure 2.

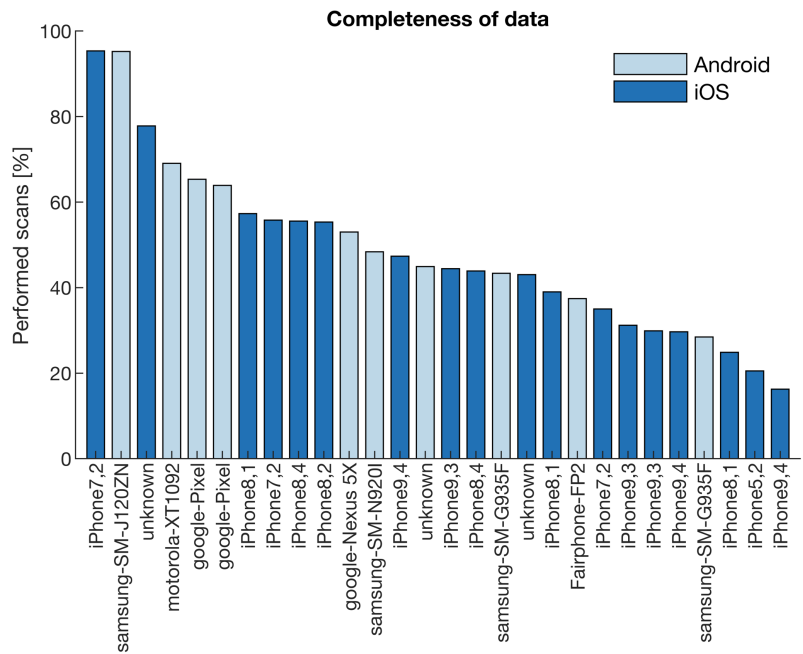

**Figure 2.** Completeness of data acquired by different devices used by participants

### Passive Data Collection

In this study, we collected two types of sensor data (Bluetooth and GPS) using the Socialise app. Both types of data may provide behavioral indicators of mental health.

#### Bluetooth Connectivity

When assessing the number of mobile phone devices that were detected using Bluetooth, we observed large variability between participants, both in the total number of devices that were detected and the ratio of known and unknown devices, as seen in the top panel of Figure 3. When considering the average number of nearby mobile phones at different times of the day, few nearby devices were detected during sleeping time (0–6 am), and they were mostly known devices, as seen in the bottom panel of Figure 3. In contrast, office hours had the most device detections and also showed the highest percentage of unknown devices. In the evening, the number of known devices stabilized, whereas the number of unknown devices gradually decreased.

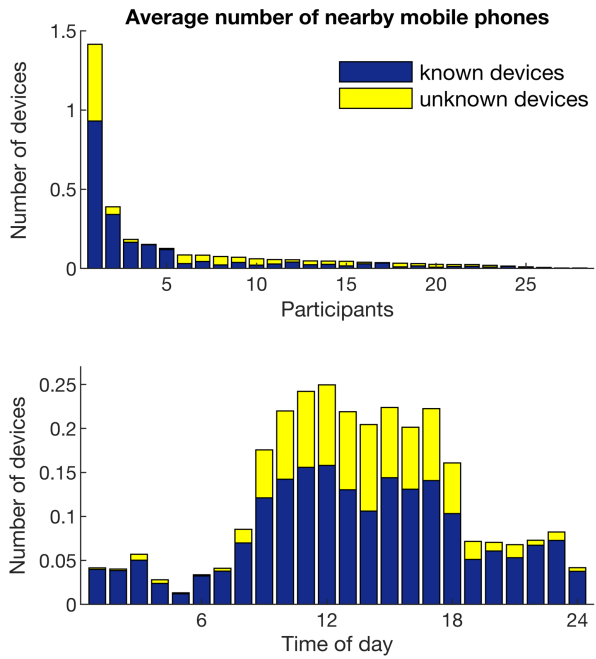

**Figure 3.** Number of Bluetooth devices that were detected. Blue indicates known devices and yellow unknown devices

*Global Positioning System-Location and Mobility*

Location data were captured from participants throughout Australia. The top panel in Figure 4 shows the locations (latitude and longitude) of participants during the 4-week study overlaid on Google maps. Data of individual participants are color coded. The number of location clusters identified for each participant ranged from 4 to 30 with a median of 8 clusters. The bottom panel of Figure 4 shows clusters extracted from a representative participant. Dots represent the centroid of different clusters and the size of dots indicates the number of samples captured within each cluster.

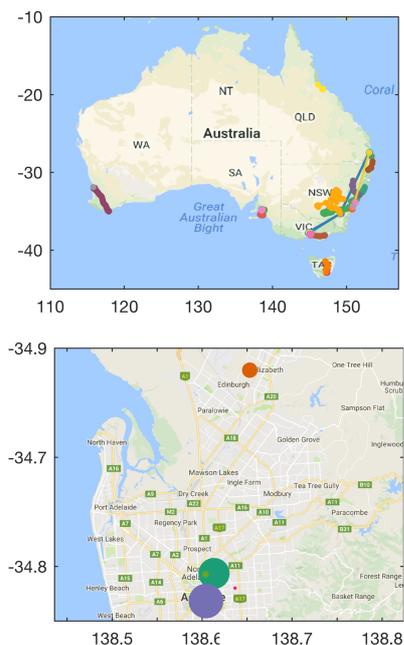

**Figure 4.** GPS location data of participants during study

Figure 5 shows the circadian movement measured at scanning intervals of 3 to 8 minutes (displayed as separate lines). Circadian movement measures to what extent the participants' sequence of locations follows a 24 to hour rhythm. Lower circadian movement scores indicate that location changes revealed a weaker 24-hour rhythm. A repeated-measures ANOVA showed no significant effect of scanning interval on circadian movement ($F_{3,69}$=2.31, $P$=.08), indicating that different scanning intervals did not introduce a significant bias in estimating circadian movement. Cronbach alpha was .79 (95% CI 0.61-0.89), indicating an acceptable consistency in the circadian movement estimated at different scanning intervals in different weeks.

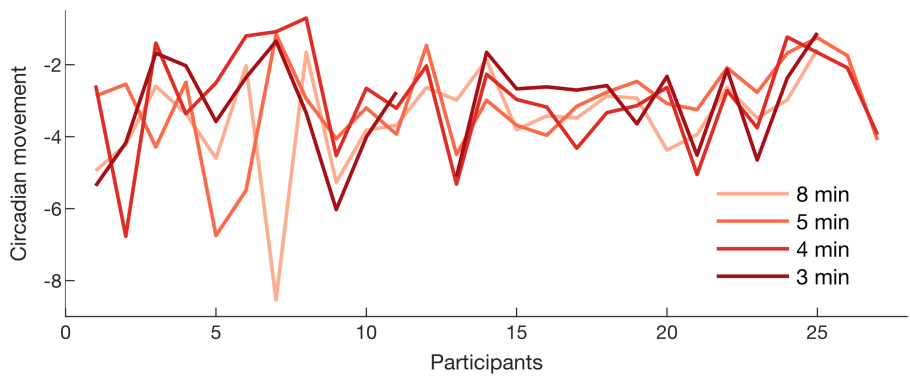

**Figure 5.** Circadian movement estimated from GPS data of individual participants

User Experience

*Battery Consumption*

We considered that users typically charge their phones once per day and are awake typically from 6 am to 10 pm (16 hours). With operation of the app, battery life should ideally last at least 16 hours after a full recharge. After systematically varying the time interval between GPS and Bluetooth scans, we used a robust fitting algorithm to estimate the average battery consumption of the Socialise app across devices and scanning rates. Based on the fitted blue regression line seen in Figure 6, we estimated that the average battery life was 21.3 hours when the app did not scan at all, and was reduced to 18.8 hours when the app scanned every 5 minutes, resulting in a reduction of 2.5 hours (12%) in battery life. Gray lines show data from individual devices, showing that scanning at 5-minute intervals permitted 5-29 hours of battery life. At this scanning rate, 13 out of 16 devices (81%) had an average battery life of more than 16 hours. At an interscan interval of 3 minutes, average battery life was further reduced to 17.4 hours. In comparison to the small reduction in battery life at increased scanning rates, we observed large variability in battery life across devices, as seen in Figure 6.

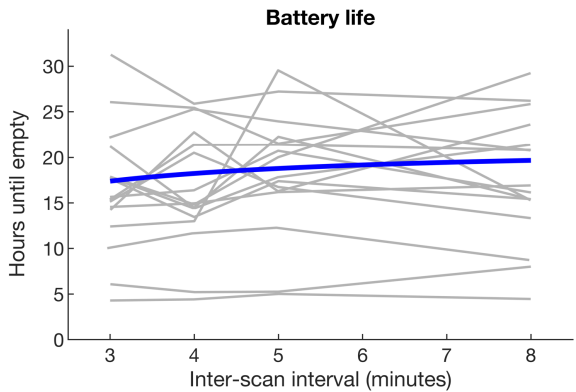

**Figure 6.** Battery life as function of the scanning rate of the Socialise app

Different scanning rates were also subjectively evaluated by asking participants whether they felt that the app impacted their mobile phone's battery life at the end of each week. Participants were asked the question "In the last week, did the app impact the battery life of your phone?" Overall, 23 participants answered the survey question, and 56 ratings were provided over the course of the study. Figure 7 shows the perceived impact of the app on battery life for different scanning frequencies. The percentage of respondents is shown for each of the scores of a 7-point Likert scale, where higher scores indicate greater impact. Colors indicate different scanning rates (once every 8, 5, 4, and 3 minutes with n=18 for 3 and 5 minutes and n=10 for 5 and 8 minutes). The majority of participants reported that battery life was affected by the Socialise app, in particular at higher scanning rates (every 3 or 4 minutes).

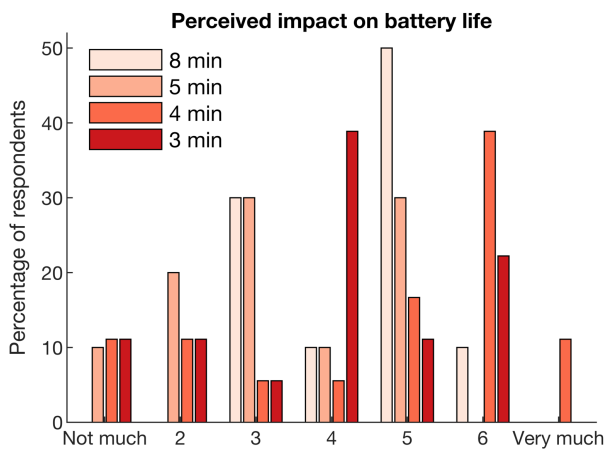

**Figure 7.** Participant ratings of the impact of the Socialise app on battery life

*Usability*

As part of an iterative design and development process, we asked participants to report any problems they experienced in using the Socialise app. Overall, 30 participants (30/32, 94%) answered questions about problems associated with installing and opening the app with half (15/30, 50%) indicating they experienced problems. The most common problem was difficulty logging into the app with the unique participant code (7 participants; Table 2). Many reported problems were technical, which are difficult to address in a preemptive manner because they often depend on user-dependent factors, such as the type, brand, and age of their mobile phones and user behavior (eg, skimming instructions).

**Table 2.** Problems experienced installing and opening the Socialise app (n=30).

| Response | n (%) | Potential solutions |
| --- | --- | --- |
| No problems | 15 (50) | — |
| Problem logging into app | 7 (23) | Simplify token |
| User self-identified lack of proficiency with technology | 2 (7) | Improve instructions |
| App not loading | 1 (3) | Improve app release |
| App needing reinstallation | 1 (3) | Improve app release |
| Phone settings blocking app | 1 (3) | Improve instructions |
| Unspecified problem | 3 (10) | — |

Fewer participants (23/32, 72%) answered questions about problems they experienced while running the app; these questions were administered at the end of each week. In total, questions were answered 56 times over the course of the study. Just under half (11/23, 48%) of the respondents reported problems running the app, and a problem was identified 32% (18/56) of the time (Table 3). The most common problem was that the app provided a notification to participants stating that they had restarted their phone when users, in fact, had not (7 times). Again, it is evident that a number of encountered problems were technical and, as before, they may be due to mobile phone and user behavior-related factors.

**Table 3.** Problems experienced running the Socialise app.

| Response | n (%)[a] | Potential solutions |
| --- | --- | --- |
| No problem | 38 (68) | — |
| App notification telling users they restarted phones | 7 (13) | Only send notification if no data are uploaded to database |
| Noticeable battery loss | 2 (4) | Reduce scanning rate |
| Difficulty sending emails after app installation | 2 (4) | — |
| App not presenting questionnaires | 1 (2) | Check scheduling function of Socialise app |
| App not scanning | 1 (2) | Check settings |
| Annoying to keep app open; accidentally swipe closed | 1 (2) | Send notification if no data are uploaded to database |
| Unsure if app running properly | 1 (2) | Improve instructions |
| Unsure about what they should be doing with app | 1 (2) | Improve instructions |
| App not working | 1 (2) | Check phone model and operating system |
| Unspecified problem | 1 (2) | — |

[a]Participants were asked to answer questions about problems running the app four times during the study. Twenty-three unique participants answered these questions, yielding 56 responses.

### Ethics

To explore ethics and privacy considerations of passive mobile phone sensor data collection, we included a set of survey questions about the acceptability of sensor data collection and some contextual information about that acceptability. Survey questions were administered at the end of the feasibility study (n=13) using a 5-point Likert scale. The top panel of Figure 8 shows that most participants expressed comfort with all aspects of data collection; 77% (10/13) of the participants were either comfortable or very comfortable with GPS, 53% (7/13) with Bluetooth, and 100% (9/9) with questionnaires. A repeated-measures ANOVA showed no main effect of data type ($F_{2,24}$=2.09, $P$=.15, n=13). We also asked participants how comfortable they were with data collection in different contexts, as seen in the bottom panel of Figure 8. Repeated-measures ANOVA showed a main effect of context ($F_{2.4,29.2}$=7.48, $P$=.01). Post hoc $t$ tests showed that participants were more comfortable with data collection for research than for advertising ($t_{12}$=−3.99, $P$=.002) and for medical intervention than for advertising ($t_{12}$=3.89, $P$=.003).

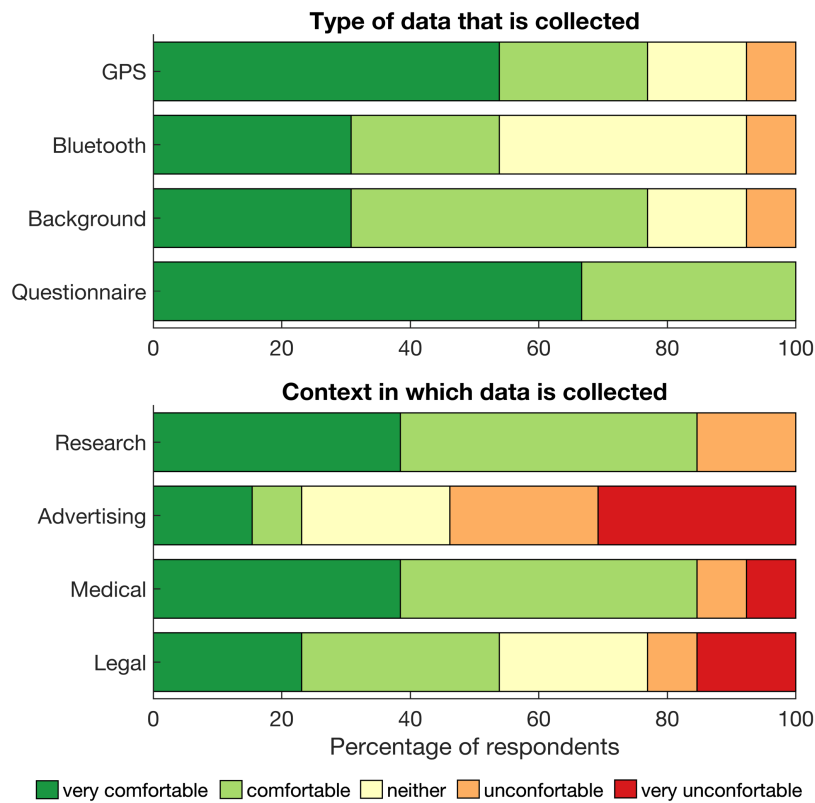

**Figure 8**. Participants' comfort with aspects and context of background data collection

Thematic analysis of responses to open questions revealed the following 3 main themes: uncertainty around the purpose of data collection, helpfulness of data donation to a respected research institute with a secondary theme of trust, and the personal impact of using the app including a secondary theme of perceived impact on privacy.

Participant 11 (henceforth P11), who said they were "Neither comfortable nor uncomfortable" with GPS data collection, explained that "[I was] ok; however, as I was not fully aware of the intentions of the collection of the GPS data and my battery life declining, I started to then get uncomfortable." Another participant, who also said "neither" for both Bluetooth and GPS tracking said, "I wasn't sure what the purpose was," and "[I] don't understand the implications of this at all" [P12]. P13 said, "Why collect this data?" and "[I] cannot see what value it would be other than to satisfy arbitrary research goals" and felt it to be "an invasion of my privacy." These responses imply that although the level of discomfort was low overall, a degree of uncertainty existed around the purpose of data collection, and this uncertainty increased discomfort.

Another theme related to the motivation of being helpful to the research or the Institute by providing data. Overall, 4 of the 13 respondents mentioned being helpful as a motivation. P3 was "very comfortable" with GPS tracking and said, "[I] wanted to help in some way." P2 was quite comfortable with the app running in the background "because I realize that information will be used for the betterment of [the] community." P7 said, "[I] would like to do anything I can that might help more study," and P8 would continue using the app or "anything that could help." This theme is unsurprising given that these users are on a volunteer research register. A second and related theme was around trust. One user explained, "[I] trust the Black Dog [Institute]" (P3) and was therefore comfortable with passive data collection.

Many participants framed their level of comfort with data collection in terms of its perceived effect or impact on them. One participant was "very comfortable" with GPS tracking because "it didn't affect me" (P4). Others said, "[it] does not bother me" (P2), "[it] did not bother me" (P10), or "[I] did not think much about it" (P9). However, another user who said, "[I was] comfortable" with GPS data collection, explained: "I actually forgot most of the time that it was collecting it. Which slightly made me uncomfortable just in regard to how easily it can happen" (P5). P11, who answered "neither" for effect or impact, said that GPS tracking was impacted by what... was draining their battery. P2 also said, "Bluetooth drains battery" and "[I was] uncomfortable" with the Bluetooth being on, but also that it was "not a huge problem." Finally, one user was "uncomfortable" with GPS tracking, explaining, "I believe it is an invasion of my privacy" (P13). However, the same user believed there were "no privacy issues" with Bluetooth data collection.

Another aspect of impact on users was the idea of perceived benefit or lack thereof for them. When responding to a question about whether they would continue to the use the app: "If the app were to be modified showing people you meet and giving information about what it means, I probably would [continue using it]" (P1). However, others said they "don't see a use for it" (P5) and "[were] not sure how useful it would be for me" (P9). This is not surprising considering that the app is solely for data collection. However, it shows that participants would expect to receive information that they can interpret themselves.

## Discussion

### Principle Findings

A feasibility study was conducted to test the Socialise app and examine challenges of using mobile phone sensor technology for mental health research. Sensor data (Bluetooth, GPS, and battery status) was collected for 4 weeks, and views of participants about acceptability of passive sensor technology were investigated. We were able to collect sensor data for about half of the scheduled scans. Social context, location clusters, and circadian movement were features extracted from sensor data to examine behavioral markers that can be obtained using the app. Battery life was reduced by 2.5 hours when scanning every 5 minutes. Despite this limited impact on battery life, most participants reported that running the app noticeably affected their battery life. Participants reported the purpose of data collection, trust in the organization that collects data, and perceived impact on privacy as important considerations for acceptability of passive data collection.

### Behavioral Markers

Instead of assessing social connections between participants, Bluetooth data were used to make a coarse estimate of human density around the participant, which provides a rough proxy for social context. The number and familiarity of devices detected were used to differentiate social contexts. Specifically, more unfamiliar devices were detected during work hours, and fewer familiar devices were detected in the evening. This pattern largely matched that observed by Do et al [47], although the number of overall devices that were detected in our study was lower. This may be partly because we recorded only Bluetooth data from mobile phone devices while filtering out other Bluetooth devices.

We extracted two features from GPS data previously shown to have strong association with self-reported mental health data [7]: circadian movement and location clusters. Circadian movement measures to what extent participants' sequence of locations follows a 24-hour rhythm. Comparing circadian movement assessed separately each week to values across weeks revealed good reliability (Cronbach alpha .79), indicating acceptable consistency in circadian movement estimated in different weeks at different scanning rates. Circadian movement was estimated over 1 week of GPS data, and consistency may be further improved by estimating circadian movement over longer time intervals. We also used a clustering algorithm to identify the number of location clusters that each participant visited. The number of clusters ranged from 4-30 with a median of 8 clusters, which was higher than the number of location clusters reported by Saeb et al [7], ranging from 1-9 with an average of 4.1 clusters. This may be partly due to geographical differences between studies (Australia vs United States). Human mobility patterns are strongly shaped by demographic parameters and geographical contexts, such as age and population density, and it should therefore be determined whether behavioral markers extracted from GPS data are universal or context-dependent [50,51].

### Technical Challenges

We were able to collect sensor data for about half of the scheduled scans (Android 55%, iOS 45%). The Socialise app (v0.2) incorporated two technical modifications (ie, using push notifications to trigger scans and using significant-change location service to alert participants when their phone restarted and remind them to resume data collection) to improve data completeness on iOS devices compared with our previous studies, which revealed significant disparity between Android and iOS data acquisition rates using previous versions of the app [43,44]. The 50% data rate in this study is similar to the rate reported in a study using Purple Robot, in which 28 of 40 participants (70%) had data available for more than 50% of the time [7]. However, GPS data of only 18 participants (45%) were used for location analysis in that study, suggesting that the GPS data rate may have been lower. Likewise, in a study using Beiwe in a cohort with schizophrenia, the mean coverage of GPS and accelerometer data were 50% and 47%, respectively [52]. Missing data may limit the number of participants for whom features can be reliably estimated and may also introduce bias in outcome measures extracted from sensor data, for example, participants with fewer data points will appear to have fewer social connections [53]. Interestingly, a recent pilot study (N=16) found that the total coverage of sensor data is itself associated with self-reported clinical symptoms [52].

We found that the Socialise app, when scanning every 5 minutes, reduced battery life from 21.3 hours to 18.8 hours, a 12% reduction. We used silent push notifications to trigger scans intermittently because continuously sampling sensor data would drain the phone's battery in a few hours. Pendão et al [54] estimated that GPS consumed 7% and Bluetooth consumed 4% of total battery power per hour when sampling continuously or 1% and 3%, respectively, when sampling periodically. Therefore, a straightforward solution to conserve battery life is to adjust intervals between data collection points. Longer time intervals between scans and shorter scanning durations can reduce battery consumption, but scanning durations that are too short may not yield meaningful sensor information [23]. Although we used silent push notifications to schedule intermittent scans, other apps use an alternating on-cycle to off-cycle schedule, in which GPS was scheduled to collect data with 1 Hz frequency for a 60-seconds on-cycle, followed by a 600-seconds off-cycle [52]. Another approach to conserve battery is to use conditional sensor activation, for example, adaptive energy allocation [55] and hierarchical sensor management [23]. These solutions reduce the activation of specific sensors at times when they are not needed.

### Ethical Considerations

The collection of sensor data involves large quantities of individualized social and behavioral data, and security and privacy have been recognized as a high priority [9,10]. Our participants reported that the purpose of data collection was an important consideration to weigh against any perceived privacy risks, which relates to the theme of uncertainty around purposes of data collection. The consent process for mental health data collection is therefore of importance with regard to both articulating this purpose and outlining confidentiality and risk of harm to patients [35]. Patient safety should be built into the design of data collection apps. Although this study did not collect mental health data, we intend to use the Socialise app in future studies to assess the mental health symptoms of participants. As such, we have built into the Socialise app a safety alert system, by which participants who indicate high scores on mental health questionnaires will be immediately given contact information about support services and be contacted by a mental health professional to provide additional support. This is consistent with the views of practitioners who have emphasized the importance of including contacts for medical professionals or other services in case of emergency or the need for immediate help [9]. Patients should be made aware of the standard turnaround time for a response to requests for help [2] and administering organizations should ensure that these expectations are clearly defined and consistently met [2].

Our results revealed a degree of uncertainty about the purpose of the study, suggesting that many participants took part without necessarily feeling informed about reasons for it. The communication of purpose should therefore be improved for future studies. Hogle [56] emphasized the need to make a clear distinction whether health-related data are collected for population-level research as opposed to individual, personal treatment or identification of issues. In addition, data processing techniques are often opaque to users, and informed consent may thus be difficult to achieve [42]. Respondents also emphasized their willingness to help the organization with its research and their trust in the organization as a stand-in for certainty about how data would be used. We believe that researchers should not rely on organizational trust as a stand-in for true understanding and informed consent because there is a risk of breach of trust if data are not used as expected.

Other issues included data ownership and the direction of any benefits created, considering that the data are from users [40]. Pentland et al [57] argued that participants should have ownership over their own data, by which they mean that app users should maintain the rights of possession, use, and disposal with some limitations on the right to disclose data about others in one's network. This can be achieved by holding users' data much as a bank would, with informed consent, or by storing data locally on a user's device and requiring upload for analysis [57]. However, when it comes to data, it is those with the capacity to store, analyze, and transfer data who have meaningful power over it; therefore, the concept of data ownership is limited [58].

Passive sensor data may be used for predictive analytics to identify those at risk of mental health issues. However, there is a possibility that predictive models may increase inequalities for vulnerable groups [40], particularly when commercial interests are at play. Psychiatric profiling will identify some as being at high risk, which may shape self-perception [59] and beliefs about an individual. This is particularly significant if the individual is a minor [2]. Hence, nonmedical and commercial use of this data to estimate mental state and behavior is an area of concern [2].

## Recommendations

Based on these findings and the literature on passive sensing, usability, and ethics, we make the following recommendations for future research on passive sensing in mental health.

### *Reporting of Data Completeness and Battery Consumption to Benchmark Different Technical Solutions*

Standard reporting of meta-data will enable benchmarking of apps and identification of technical obstacles and solutions for sensor data collection across devices and operating systems. For example, we estimated that the Socialise app reduced battery life by 2.5 hours when scanning every 5 minutes. Although the app had small effect on battery consumption (81% of devices had an average battery life of more than 16 hours), users were very sensitive to battery performance. Standard reporting of data rates and battery consumption will allow quantitative comparisons between approaches and develop technical solutions that meet user expectations on battery life.

### *Releasing Source Code of Data Acquisition Platforms and Feature Extraction Methods*

The number of mobile phone apps for passive sensing is still increasing, but differences in methodology and feature extraction methods can impede the reproducibility of findings. This can be overcome with a commitment to open science because a number of elements of passive data research could be shared. Currently, several sensing platforms are open source, such as Purple Robot [6] and recently, Beiwe [52]. Following this lead, methods for feature extraction could be made open source, such that scripts are available for use on different data sources, providing consistency in feature extraction. Finally, the data itself should be made available on open data repositories to enable data aggregation across studies to test potential markers in larger samples, resulting in more reproducible results [60]. However, data sharing not only has great potential but also involves concerns about privacy, confidentiality, and control of data on individuals [61]. These concerns particularly apply to sensor data such as GPS that can be reidentified [62]. Databases that allow analysis to be conducted without access to raw data may be one potential solution.

*Identifying a Limited Number of Key Markers for Mental Health*

Although the use of passive data in mental health is still exploratory, researchers need to move toward agreement on best practice methods and features. The current unrestricted number of features has the danger of inflating degrees of freedom and may endanger replicability of findings [63]. Practices such as preregistration of study hypotheses and proposed methods to quantify features could help reduce spurious correlations and will be key in identifying reliable markers of mental health [64]. However, work with different sensor modalities is at different stages of development. For example, a number of GPS features have been identified and replicated [6], whereas potential markers of social connectedness using Bluetooth data still require research to assess predictive value in open network settings. This development of new methods of data analysis is indeed one of the most immediate challenges [5]. Once candidate methods have been identified, and it will be important to test these markers in larger longitudinal studies to see whether they predict the development of mental health problems and can be used to support prevention and early intervention programs [65].

*Providing Meaningful Feedback to Users*

User engagement is also a key requirement for successful implementation of sensor technology in mental health research. Investigating user experience can help us understand user expectations and improve user engagement and retention [66]. Although passive data collection is designed to be unobtrusive, perceived benefit is an important consideration for continued use of mental health apps. A user-centric design process [27] and the American Psychiatric Association's app evaluation model [67] should be followed to provide meaningful user feedback from sensor data. We also recommend using more robust measures for informed consent, considering the opacity of data analysis techniques and purposes [47] and engaging users with informative feedback derived from their data.

*Transparency in the Purpose of Data Collection*

Evidence from the literature and participant responses suggests that purposes of data collection are important as well as the awareness of the user. The use of data was found to be most the important factor in a person's willingness to share their electronic personal health data [10], and participants cared most about the specific purpose for using their health information [68]. Rothstein argued that there is too much emphasis on privacy when the concern should be about autonomy [69]. This refers to the informed consent process, during which researchers should ensure understanding and enable autonomous and active consent on that basis [69]. It is therefore recommended that researchers take care to ensure that the consent process allows participants really to understand the purpose of the research. This, in turn, is likely to increase the level of comfort with data collection.

## Conclusion

The use of passive data in mental health research has the potential to change the nature of identification and treatment of mental health disorders. Early identification of behavioral markers of mental health problems will allow us to preempt rather than respond, and understanding idiosyncratic patterns will enable personalized dynamic treatment delivered at the moment. Although a number of significant technological and broader challenges exist, we believe that open science, user involvement, collaborative partnerships, and transparency in our attempts, successes, and failures will bring us closer to this goal.

## Acknowledgments

This research was financially supported by the National Health and Medical Research Council (NHMRC) Centre of Research Excellence in Suicide Prevention APP1042580 and NHMRC John Cade Fellowship APP1056964. Tjeerd Boonstra was supported by a NARSAD Young Investigator Grant from the Brain & Behavior Research Foundation.

## Conflicts of Interest

None declared.

## Data Availability

Data used in this study is available at Zenodo: http://doi.org/10.5281/zenodo.1238226. One participant did not consent to have individual data made publicly available. We did not share GPS data because this would allow reidentification of participants. The Matlab scripts used to analyze data are available at Zenodo: http://doi.org/10.5281/zenodo.1238408.

## Abbreviations

ANOVA: analysis of variance

EMA: ecological momentary assessment

GPS: global positioning system

iOS: iPhone OS

NHMRC: National Health and Medical Research Council

PTSD: Post-Traumatic Stress Disorder